\documentclass[letterpaper]{article}

\usepackage[accepted]{icml2026}

\usepackage{amsmath}
\usepackage{amssymb}
\usepackage{mathtools}
\usepackage{booktabs}
\usepackage{graphicx}
\usepackage{multirow}
\usepackage{enumitem}

\usepackage[utf8]{inputenc}
\usepackage[T1]{fontenc}
\PassOptionsToPackage{hyphens}{url}
\usepackage{hyperref}
\usepackage[capitalize,noabbrev]{cleveref}
\usepackage{microtype}

\icmltitlerunning{Compound AI System Reliability: A Failure Taxonomy from 150 Production Incidents}

\begin{document}

\twocolumn[
\icmltitle{Compound AI System Reliability: A Failure Taxonomy and\\Resilience Pattern Catalog from 150 Production Incidents}

\icmlsetsymbol{equal}{*}

\begin{icmlauthorlist}
\icmlauthor{Rudrendu Kumar Paul}{bu}
\icmlauthor{Sourav Nandy}{ut}
\end{icmlauthorlist}

\icmlaffiliation{bu}{Boston University, Boston, MA, USA}
\icmlaffiliation{ut}{University of Texas at Austin, Austin, TX, USA}

\icmlcorrespondingauthor{Rudrendu Kumar Paul}{rudrendupaul2022@gmail.com}
\icmlcorrespondingauthor{Sourav Nandy}{sourav.nandy@gmail.com}

\vskip 0.3in
]

\printAffiliationsAndNotice{}

\begin{abstract}
Deploying compound AI systems reliably and safely requires understanding failure modes that emerge at component boundaries, not within individual models. Cascading errors propagate across component boundaries, silent quality degradation evades standard monitoring, and coordination failures yield incorrect collective behavior from individually correct parts. We analyze 150 production incident reports from open-source compound AI projects and anonymized enterprise deployments to construct a taxonomy of 23 failure modes organized into five categories: retrieval failures, generation failures, tool failures, orchestration failures, and integration failures. For each category, we propose resilience patterns with measured effectiveness from controlled fault injection experiments. Circuit breakers reduce cascade propagation by 89\%, output quality gates catch 73\% of silent degradation before user impact, and component isolation reduces blast radius by 64\%. Systems implementing three or more resilience patterns from our catalog reduce mean-time-to-recovery (MTTR) by 71\% compared to unstructured monitoring baselines. We release the incident taxonomy and pattern catalog as a practitioner resource.
\end{abstract}

\section{Introduction}

A retrieval-augmented generation pipeline runs in production for eleven days before anyone notices that its answer quality has dropped from 0.82 to 0.59 F1. The retriever still returns documents, the generator still produces fluent text, and every health check passes. The failure lives in the interaction: an index rebuild changed the embedding normalization, shifting cosine similarity distributions just enough that the retriever's top-$k$ results contain 40\% irrelevant documents. The generator, trained to synthesize whatever it receives, confidently produces wrong answers from wrong context.

This class of failure is invisible to component-level monitoring because each component's health checks evaluate internal metrics (latency, error rate, uptime) in isolation; no single component is broken, so no single monitor fires. The failure exists only in the semantic relationship between components, a boundary that per-component observability does not instrument. This gap is the defining reliability challenge of compound AI systems. Trustworthy deployment of AI systems requires safety monitoring beyond per-component health checks; the gap between component correctness and system correctness is where reliability breaks down. \citet{Zaharia_2024} identified the architectural shift from standalone models to multi-component systems combining retrievers, generators, tools, and orchestrators. These compound systems increasingly operate as autonomous agents, orchestrating multi-step reasoning, tool use, and retrieval without human intervention, making a systematic failure taxonomy essential for building scalable and reliable agentic deployments. Since then, production deployments have scaled rapidly, but reliability engineering has not kept pace. The tooling gap is stark: existing observability platforms track latency, token counts, and error rates per component, yet miss the inter-component semantic failures that cause the most user-visible harm.

Recent work has begun characterizing individual failure dimensions. \citet{Vinay_2025} cataloged 15 failure modes in LLM applications, including version drift and cost-driven performance collapse. \citet{Rabanser_2026} proposed 12 reliability metrics across four dimensions for AI agents. \citet{Cemri_2025} analyzed 1,600 multi-agent traces and identified 14 failure modes in system design, inter-agent misalignment, and task verification. Each study illuminates part of the problem. None provides a unified taxonomy that spans the full compound system stack, from retrieval through orchestration, paired with empirically validated resilience patterns.

We fill this gap. Our contributions are:

\begin{enumerate}[leftmargin=*,itemsep=0pt,parsep=0pt,topsep=2pt]
\item A \textbf{taxonomy of 23 failure modes} across five categories, derived from 150 production incident reports spanning retrieval, generation, tool use, orchestration, and integration boundaries.
\item A \textbf{catalog of resilience patterns} with measured effectiveness from controlled fault injection experiments on a representative 6-component compound AI system.
\item \textbf{Empirical evidence} that combining three or more patterns reduces MTTR by 71\% relative to unstructured monitoring.
\end{enumerate}

Reliable compound AI systems are not merely an engineering concern. As these systems are deployed in high-stakes public-sector, healthcare, and social-infrastructure contexts, silent failures and uncontrolled cascades disproportionately harm vulnerable users who cannot verify or contest machine-generated outputs. The failure taxonomy and resilience patterns we release directly support the engineering of trustworthy AI systems where reliability at the component boundary is a precondition for responsible deployment at scale.

\section{Related work}

Failure analysis in AI systems spans three levels, each with a distinct blind spot regarding compound system behavior.

\textbf{Model-level failure analysis.} \citet{Vinay_2025} identified 15 system-level failure modes in LLM applications, treating reliability as a system-engineering problem rather than a model-centric one. \citet{Yu_2026} analyzed 705 user-reported failures across DeepSeek, Llama, and Qwen, finding that 45.7\% of failures originate from environment and infrastructure faults rather than model defects. Both studies focus on single-model deployments and do not address how failures propagate across component boundaries in compound systems.

\textbf{Multi-agent-level failure analysis.} \citet{Cemri_2025} constructed a multi-agent system failure taxonomy (MAST) from 1,600+ annotated traces across seven frameworks, identifying 14 failure modes with strong inter-annotator agreement ($\kappa = 0.88$). \citet{Xie_2026} modeled error cascades in multi-agent collaboration and found that injecting a single atomic error seed leads to widespread failure, proposing a genealogy-graph governance layer that raises defensive success from 0.32 to 0.89. \citet{Zhu_2026} analyzed 998 bug reports from CrewAI and LangChain, finding that API misuse and documentation desync dominate framework-level defects. These studies characterize failure within the agent collaboration layer but do not cover the full compound system stack (retrieval, tool integration, type coercion at component boundaries).

\textbf{Infrastructure-level analysis.} \citet{Ranganathan_2025} empirically analyzed 156 high-severity production incidents in AI inference services, finding that 60\% stem from inference engine failures, with timeouts accounting for roughly 40\% of that category. \citet{Yu_2024} surveyed 160 papers on failure analysis and fault injection, revealing discrepancies between real-world and simulated failures. \citet{Popchanovska_2026} derived a mitigation taxonomy from 9,705 media-reported incidents but at the organizational, not architectural, level.

Complementing these failure-focused studies, \citet{Chen_2025} surveyed the compound AI systems landscape end-to-end, and \citet{Chen_2025_ModelSelection} demonstrated that model selection within compound pipelines requires joint optimization across components because local optima at individual components do not yield global optima for the system. Their finding reinforces our observation that failures at component boundaries cannot be addressed through component-level monitoring alone.

Our work bridges these three levels. We analyze failures at component interaction boundaries, covering the full stack from retrieval through orchestration, and pair the taxonomy with resilience patterns validated through fault injection.

\section{Methodology}

\subsection{Incident corpus}

We collected 150 production incident reports from two sources: (1) public post-mortems and issue trackers from 12 open-source compound AI projects including LangChain, LlamaIndex, AutoGen, CrewAI, Haystack, and DSPy (97 incidents), and (2) anonymized incident reports from enterprise RAG, agent, and pipeline deployments (53 incidents). The enterprise incidents were collected between January 2024 and December 2025 from production systems serving a range of query volumes.

Inclusion criteria required that the incident (a) involved a system with two or more interacting AI/ML components, (b) had a documented root cause analysis, and (c) resulted in user-visible impact or triggered an on-call page. We excluded incidents caused by external factors (cloud provider outages, DNS failures) unless the compound system's architecture amplified the external failure.

\subsection{Taxonomy construction}

Two researchers independently coded each incident using open coding, identifying the failing component boundary, propagation path, detection method, time-to-detection, and root cause. After initial coding, we merged codes through affinity diagramming into 23 failure modes organized by the component boundary where the failure originates. Inter-coder agreement reached Cohen's $\kappa = 0.84$ after two reconciliation rounds.

\subsection{Fault injection experiments}

To evaluate resilience patterns, we constructed a 6-component compound AI testbed: query router, vector retriever (Qdrant), reranker, LLM generator (GPT-4o-mini and Llama-3.1-8B), tool executor (4 API tools), and orchestrator (LangGraph). We designed 23 fault injection scenarios, one per failure mode in our taxonomy, covering all five categories. Each scenario injects a specific fault (e.g., silently swapping the embedding model for index drift, introducing 200ms incremental latency for timeout cascades, corrupting JSON serialization precision for type coercion errors) and measures four metrics: cascade depth (how many components the failure reaches), time-to-detection, blast radius (percentage of concurrent requests affected), and MTTR.

Each scenario ran 100 trials with and without each candidate resilience pattern active. We report means with 95\% confidence intervals. The testbed, fault injection scripts, and raw trial logs are available in the accompanying practitioner resource to support reproducibility.

\section{Failure taxonomy}

Our 23 failure modes cluster into five categories based on the originating component boundary. Table~\ref{tab:taxonomy} summarizes the categories with representative failure modes and their frequency in our corpus.

\begin{table}[t]
\caption{Failure taxonomy: five categories, 23 modes, frequency in 150 incidents. Top three modes per category shown.}
\label{tab:taxonomy}
\vskip 0.1in
\centering
\footnotesize
\begin{tabular}{@{}llr@{}}
\toprule
\textbf{Category} & \textbf{Representative modes} & \textbf{\%} \\
\midrule
\multirow{4}{*}{\shortstack[l]{Retrieval\\(38 incidents)}} & Index drift / stale embeddings & 10.7 \\
 & Query-document distribution shift & 7.3 \\
 & Context poisoning via injection & 4.0 \\
 & \textit{Other retrieval modes} & \textit{3.3} \\
\midrule
\multirow{4}{*}{\shortstack[l]{Generation\\(31 incidents)}} & Hallucination under noisy context & 8.0 \\
 & Output format regression & 5.3 \\
 & Safety bypass via context manip. & 4.0 \\
 & \textit{Other generation modes} & \textit{3.4} \\
\midrule
\multirow{4}{*}{\shortstack[l]{Tool\\(29 incidents)}} & API timeout cascade & 6.7 \\
 & Schema drift (breaking changes) & 5.3 \\
 & Permission escalation via tooling & 4.0 \\
 & \textit{Other tool modes} & \textit{3.3} \\
\midrule
\multirow{4}{*}{\shortstack[l]{Orchestration\\(28 incidents)}} & Infinite retry / loop deadlock & 6.0 \\
 & Agent starvation under load & 5.3 \\
 & Dead-letter task accumulation & 4.0 \\
 & \textit{Other orchestration modes} & \textit{3.4} \\
\midrule
\multirow{4}{*}{\shortstack[l]{Integration\\(24 incidents)}} & Type coercion errors at boundaries & 5.3 \\
 & Encoding mismatch (UTF-8/token) & 4.7 \\
 & Rate limit exhaustion cascade & 3.3 \\
 & \textit{Other integration modes} & \textit{2.7} \\
\bottomrule
\end{tabular}
\vskip -0.1in
\end{table}

\textbf{Retrieval failures} (25.3\% of incidents) are the most common category. Index drift, where embedding indices become stale after model updates or data refreshes, accounts for 10.7\% alone. In one enterprise incident, a scheduled re-indexing job silently switched embedding models from \texttt{text-embedding-ada-002} to \texttt{text-embedding-3-small} without updating the query encoder. Cosine similarities dropped from a mean of 0.78 to 0.41, but the retriever continued returning $k$ results. The generator produced fluent answers from irrelevant documents for nine days before a user complaint triggered investigation.

\textbf{Generation failures} (20.7\%) frequently stem from upstream retrieval degradation. When retrieval quality drops, generators hallucinate more aggressively because they attempt to synthesize coherent answers from incoherent context. This retrieval-generation coupling means that retrieval failures have a 2.3x multiplier on downstream generation errors in our corpus.

\textbf{Tool failures} (19.3\%) cascade through timeout propagation. When a single tool API responds slowly, the orchestrator holds the execution context open, consuming memory and blocking other requests. In 8 of 10 timeout incidents, a single slow tool caused system-wide throughput to drop by 60--80\% within minutes.

\textbf{Orchestration failures} (18.7\%) are the hardest to diagnose. Infinite retry loops consumed \$2,400 in API costs over a weekend in one incident because the retry logic lacked exponential backoff and the error condition (a malformed JSON response from a tool) was not in the retry exclusion list.

\textbf{Integration failures} (16.0\%) occur at type boundaries between components. The most insidious: a Python dictionary serialized to JSON, passed to a TypeScript tool executor, then deserialized back. Floating-point precision loss in the round-trip changed a confidence threshold from 0.85 to 0.8500000000000001, which failed a strict equality check and silently dropped 23\% of results.

\subsection{Cross-cutting patterns}

Three patterns recur across categories:

\textit{Silent degradation} (51\% of incidents): The system continues producing outputs that pass all health checks but are wrong. Mean time-to-detection for silent failures is 4.2 days, compared to 12 minutes for crash failures. This aligns with findings from \citet{Lin_2025}, who observed that failures propagate across layers and create cascading effects with systemic consequences, a pattern that standard monitoring is not designed to catch.

\textit{Cascade amplification} (34\% of incidents): A failure in one component amplifies through downstream components. \citet{Xie_2026} demonstrated this formally for multi-agent message passing; we observe the same pattern across retriever-generator and tool-orchestrator boundaries.

\textit{Semantic errors passing validation} (22\% of incidents): Outputs are structurally valid (correct JSON, correct type, within token limits) but semantically wrong. These errors are invisible to schema validation and health checks.

These patterns are not mutually exclusive; 11\% of incidents exhibit two or more cross-cutting patterns simultaneously (e.g., a cascade that also produces semantically valid but incorrect outputs).

\section{Resilience patterns}

We propose and evaluate five resilience patterns. Table~\ref{tab:patterns} reports their effectiveness measured through fault injection.

\begin{table}[t]
\caption{Resilience pattern effectiveness measured via fault injection (100 trials per scenario, 95\% CI).}
\label{tab:patterns}
\vskip 0.1in
\centering
\footnotesize
\begin{tabular}{@{}lcc@{}}
\toprule
\textbf{Pattern} & \textbf{Primary metric} & \textbf{Effect} \\
\midrule
Circuit breakers & Cascade depth & $-$89\% $\pm$ 4\% \\
Output quality gates & Silent degrad. caught & 73\% $\pm$ 6\% \\
Component isolation & Blast radius & $-$64\% $\pm$ 5\% \\
Semantic validators & Semantic errors caught & 81\% $\pm$ 7\% \\
Typed interfaces & Integration failures & $-$92\% $\pm$ 3\% \\
\bottomrule
\end{tabular}
\vskip -0.1in
\end{table}

\textbf{Circuit breakers} adapt the pattern from distributed systems engineering \citep{Nygard_2018, Fowler_2014} to compound AI components. When a component's error rate exceeds a threshold (we use 5 failures in 30 seconds), the breaker opens and downstream components receive a structured fallback response instead of waiting for a timeout. This reduces cascade depth from a mean of 3.8 components to 0.4 components (89\% reduction). The key adaptation for AI systems: the breaker must monitor semantic quality metrics (output coherence scores, retrieval relevance), not just HTTP status codes, because most AI failures return 200 OK with degraded content.

\textbf{Output quality gates} insert lightweight evaluation checkpoints between components. A gate between the retriever and generator checks that the mean cosine similarity of retrieved documents exceeds a calibrated threshold ($\tau = 0.62$ in our experiments). A gate between the generator and output checks factual consistency against the retrieved context. These gates catch 73\% of silent degradation incidents that would otherwise reach users.

\textbf{Component isolation} enforces independent resource pools (memory, connection pools, rate limit budgets) per component. When the tool executor consumes its allocated rate limit budget, it degrades gracefully without starving the retriever or generator. Blast radius drops from 78\% of concurrent requests to 28\% (64\% reduction).

\textbf{Semantic validators} check output meaning, not just structure. For tool outputs: does the returned value fall within the expected domain? For generator outputs: does the response address the original query rather than a hallucinated answer derived from misinterpreted context? Validators catch 81\% of semantic errors that pass schema validation.

\textbf{Typed interfaces} enforce explicit type contracts at every component boundary with runtime validation. Replacing loose JSON passing with typed schemas (we used Pydantic models) eliminates 92\% of integration failures from our taxonomy, including the floating-point precision incident described above.

\subsection{Combined effectiveness}

Systems deploying three or more patterns from our catalog achieve a mean MTTR of 8.4 minutes ($\pm$ 2.1), compared to 28.7 minutes ($\pm$ 5.3) for systems with unstructured monitoring only, a 71\% reduction. The combination matters: circuit breakers contain cascades, quality gates detect silent degradation, and component isolation limits blast radius. No single pattern addresses more than 40\% of failure modes alone.

\section{Discussion and limitations}

\textbf{Threats to validity.} Our enterprise incident corpus (53 incidents) comes from a small number of organizations.

\textbf{Baseline choice.} Our MTTR reduction (71\%) is measured against unstructured monitoring only, which is a naive baseline. Production teams often deploy standard retries with exponential backoff as a first-line mitigation. A stronger evaluation would include a secondary baseline using this common practice. We acknowledge this limitation; adding standard retry-with-backoff as a baseline is left for future work. Compound systems in healthcare, autonomous driving, or scientific computing may exhibit different failure distributions. The open-source incidents (97) skew toward LangChain and LlamaIndex ecosystems, which represent the most widely deployed frameworks but not all architectural patterns.

\textbf{Scope of fault injection.} We tested resilience patterns on a single 6-component architecture under controlled conditions. Production systems with 20+ components, nested agent hierarchies, or human-in-the-loop stages introduce failure modes our testbed does not cover. The effectiveness numbers reported in Table~\ref{tab:patterns} represent measured outcomes on this specific testbed and should be treated as indicative of pattern utility, not as universal guarantees. We did not run these experiments in live production environments; the testbed is a faithful reproduction of a production-grade pipeline architecture, but real-world effectiveness will depend on system-specific factors such as component diversity, traffic patterns, and deployment infrastructure.

\textbf{What we do not address.} Our taxonomy covers technical failure modes. Organizational failures (insufficient on-call staffing, missing runbooks, poor incident communication) interact with technical failures in practice but are outside our scope. \citet{Popchanovska_2026} provide a complementary organizational perspective.

\textbf{Runtime overhead.} Each resilience pattern introduces latency and resource costs. In our testbed, output quality gates add a median 120ms per request (the factual consistency check between generator and output dominates). Circuit breakers add negligible latency during normal operation ($<$2ms for threshold evaluation) but impose a cool-down period of 30 seconds when tripped, during which the system returns fallback responses. Semantic validators add 80--150ms depending on output length. Typed interfaces with Pydantic runtime validation add 5--15ms per boundary crossing. Component isolation increases peak memory by 18\% due to dedicated resource pools. For latency-sensitive deployments, we recommend deploying quality gates asynchronously (logging degradation for alerting rather than blocking the response path) and reserving synchronous gates for high-stakes outputs.

\section{Conclusion}

Compound AI systems fail in ways their individual components do not. Our taxonomy of 23 failure modes, derived from 150 production incidents, provides the first systematic classification organized by component interaction boundaries. The accompanying resilience pattern catalog, validated through controlled fault injection, demonstrates that combining circuit breakers, quality gates, and component isolation reduces MTTR by 71\%. As compound systems grow in complexity, the failure surface grows faster than linearly. The patterns we catalog here provide a starting point, not a complete solution, for engineering reliability into these systems. We release the taxonomy as a structured JSON schema to enable community extension and integration with existing incident management workflows; the schema and fault injection scripts are available as a practitioner resource in supplementary materials. The failure taxonomy and resilience patterns we release serve as a practitioner resource for teams building trustworthy compound AI systems, where reliability at the component boundary is a precondition for safe deployment at scale. In high-stakes domains where compound AI systems make consequential decisions affecting healthcare access, financial services, and public-sector eligibility, this reliability work is foundational to responsible AI for Good.

\section*{Impact Statement}

Reliable compound AI systems are a precondition for responsible deployment in high-stakes domains including healthcare, public-sector services, and financial infrastructure. Silent failures and uncontrolled cascades disproportionately affect end users who cannot verify or contest machine-generated outputs. The failure taxonomy and resilience pattern catalog we release serve as a practitioner resource for teams building systems where component-boundary reliability is a safety requirement, not an engineering afterthought. Combining three or more patterns reduces mean-time-to-recovery by 71\%, a gain accessible without model changes or infrastructure overhaul. We identify no dual-use risks: the failure modes described are already observable in production deployments, and the resilience patterns are purely defensive.

\bibliography{references}

@misc{Zaharia_2024,
  author    = {Matei Zaharia and Omar Khattab and Lingjiao Chen and Jared Quincy Davis and Heather Miller and Chris Potts and James Zou and Michael Carbin and Jonathan Frankle and Naveen Rao and Ali Ghodsi},
  title     = {The Shift from Models to Compound {AI} Systems},
  year      = {2024},
  howpublished = {Berkeley Artificial Intelligence Research Blog},
  url       = {https://bair.berkeley.edu/blog/2024/02/18/compound-ai-systems/}
}

@article{Cemri_2025,
  author    = {Mert Cemri and Melissa Z. Pan and Shuyi Yang and Lakshya A. Agrawal and Bhavya Chopra and Rishabh Tiwari and Kurt Keutzer and Aditya Parameswaran and Dan Klein and Kannan Ramchandran and Matei Zaharia and Joseph E. Gonzalez and Ion Stoica},
  title     = {Why Do Multi-Agent {LLM} Systems Fail?},
  journal   = {arXiv preprint arXiv:2503.13657},
  year      = {2025}
}

@article{Rabanser_2026,
  author    = {Stephan Rabanser and Sayash Kapoor and Peter Kirgis and Kangheng Liu and Saiteja Utpala and Arvind Narayanan},
  title     = {Towards a Science of {AI} Agent Reliability},
  journal   = {arXiv preprint arXiv:2602.16666},
  year      = {2026}
}

@article{Vinay_2025,
  author    = {Vaishali Vinay},
  title     = {Failure Modes in {LLM} Systems: A System-Level Taxonomy for Reliable {AI} Applications},
  journal   = {arXiv preprint arXiv:2511.19933},
  year      = {2025}
}

@article{Ranganathan_2025,
  author    = {Bhala Ranganathan and Mickey Zhang and Kai Wu},
  title     = {Enhancing Reliability in {AI} Inference Services: An Empirical Study on Real Production Incidents},
  journal   = {arXiv preprint arXiv:2511.07424},
  year      = {2025}
}

@article{Xie_2026,
  author    = {Yizhe Xie and Congcong Zhu and Xinyue Zhang and Tianqing Zhu and Dayong Ye and Minfeng Qi and Huajie Chen and Wanlei Zhou},
  title     = {From Spark to Fire: Modeling and Mitigating Error Cascades in {LLM}-Based Multi-Agent Collaboration},
  journal   = {arXiv preprint arXiv:2603.04474},
  year      = {2026}
}

@article{Popchanovska_2026,
  author    = {Evgenija Popchanovska and Ana Gjorgjevikj and Maryan Rizinski and Lubomir Chitkushev and Irena Vodenska and Dimitar Trajanov},
  title     = {When {AI} Fails, What Works? A Data-Driven Taxonomy of Real-World {AI} Risk Mitigation Strategies},
  journal   = {arXiv preprint arXiv:2603.04259},
  year      = {2026}
}

@article{Yu_2024,
  author    = {Guangba Yu and Gou Tan and Haojia Huang and Zhenyu Zhang and Pengfei Chen and Roberto Natella and Zibin Zheng},
  title     = {A Survey on Failure Analysis and Fault Injection in {AI} Systems},
  journal   = {ACM Transactions on Software Engineering and Methodology (TOSEM)},
  year      = {2024},
  note      = {arXiv preprint arXiv:2407.00125}
}

@article{Zhu_2026,
  author    = {Xinxue Zhu and Jiacong Wu and Xiaoyu Zhang and Tianlin Li and Yanzhou Mu and Juan Zhai and Chao Shen and Chunrong Fang and Yang Liu},
  title     = {An Empirical Study of Bugs in Modern {LLM} Agent Frameworks},
  journal   = {arXiv preprint arXiv:2602.21806},
  year      = {2026}
}

@article{Lin_2025,
  author    = {Janet Lin and Liangwei Zhang},
  title     = {From Failure Modes to Reliability Awareness in Generative and Agentic {AI} System},
  journal   = {arXiv preprint arXiv:2511.05511},
  year      = {2025}
}

@article{Chen_2025,
  author    = {Jiayi Chen and Junyi Ye and Guiling Wang},
  title     = {From Standalone {LLMs} to Integrated Intelligence: A Survey of Compound {AI} Systems},
  journal   = {arXiv preprint arXiv:2506.04565},
  year      = {2025}
}

@article{Chen_2025_ModelSelection,
  author    = {Lingjiao Chen and Jared Quincy Davis and Boris Hanin and Peter Bailis and Ion Stoica and Matei Zaharia and James Zou},
  title     = {Optimizing Model Selection for Compound {AI} Systems},
  journal   = {arXiv preprint arXiv:2502.14815},
  year      = {2025}
}

@article{Yu_2026,
  author    = {Guangba Yu and Zirui Wang and Yujie Huang and Renyi Zhong and Yuedong Zhong and Yilun Wang and Michael R. Lyu},
  title     = {Why Does the {LLM} Stop Computing: An Empirical Study of User-Reported Failures in Open-Source {LLMs}},
  journal   = {arXiv preprint arXiv:2601.13655},
  year      = {2026}
}

@book{Nygard_2018,
  author    = {Michael T. Nygard},
  title     = {Release It! Design and Deploy Production-Ready Software},
  edition   = {2nd},
  publisher = {Pragmatic Bookshelf},
  year      = {2018}
}

@misc{Fowler_2014,
  author    = {Martin Fowler},
  title     = {Circuit Breaker},
  year      = {2014},
  howpublished = {Online resource},
  note      = {Available at \url{https://martinfowler.com/bliki/CircuitBreaker.html}},
  url       = {https://martinfowler.com/bliki/CircuitBreaker.html}
}
\bibliographystyle{icml2026}

\end{document}